# Scanning-probe spectroscopy of semiconductor donor molecules

I. Kuljanishvili, C. Kayis, J. F. Harrison, C. Piermarocchi, T. A. Kaplan and S. H. Tessmer
*Michigan State University*

L. N. Pfeiffer and K. W. West
*Bell Laboratories, Lucent Technologies*

Semiconductor devices continue to press into the nanoscale regime, and new applications have emerged for which the quantum properties of dopant atoms act as the functional part of the device, underscoring the necessity to probe the quantum structure of small numbers of dopant atoms in semiconductors[1-3]. Although dopant properties are well-understood with respect to bulk semiconductors, new questions arise in nanosystems. For example, the quantum energy levels of dopants will be affected by the proximity of nanometer-scale electrodes. Moreover, because shallow donors and acceptors are analogous to hydrogen atoms, experiments on small numbers of dopants have the potential to be a testing ground for fundamental questions of atomic and molecular physics, such as the maximum negative ionization of a molecule with a given number of positive ions[4,5]. Electron tunneling spectroscopy through isolated dopants has been observed in transport studies[6,7]. In addition, Geim and coworkers identified resonances due to two closely spaced donors, effectively forming donor molecules[8]. Here we present capacitance spectroscopy measurements of silicon donors in a gallium-arsenide heterostructure using a scanning probe technique[9,10]. In contrast to the work of Geim et al., our data show discernible peaks attributed to successive electrons entering the molecules. Hence this work represents the first addition spectrum measurement of dopant molecules. More generally, to the best of our knowledge, this study is the first example of single-electron capacitance spectroscopy performed directly with a scanning probe tip[9].

Our experimental technique is an extension of scanning charge accumulation imaging. Fig. 1a shows schematically the experiment. The key component is a metallic tip with an apex of radius ~50 nm; it is connected directly to a charge sensor that achieves a sensitivity of 0.01 $e/\sqrt{\text{Hz}}$[11]. For the capacitance spectroscopy measurements reported here, the tip's position is fixed (i.e. not scanned) at a distance of ~1 nm from the sample surface. We then monitor the tip's AC charge $q_{tip}$ in response to an AC excitation voltage $V_{exc}$ applied to an underlying electrode, as a function of DC bias voltage $V_{tip}$. As detailed in the methods section, if the quantum system below the tip can accommodate additional charge, the excitation voltage causes it to resonate between the system and the underlying electrode – giving rise to an enhanced capacitance, $C = q_{tip}/V_{exc}$. We employed a Si-doped GaAs heterostructure sample of exceptional quality grown by molecular beam epitaxy; the conduction band profile is shown in Fig. 1b. Fig. 1c shows an example of the energy landscape of the Si donor layer. For all measurements reported here the sample and tip were immersed in liquid helium-3 at a temperature of 290 mK.

As shown in Fig. 2a, the resulting measurement of the capacitance consistently showed three broad peaks in the vicinity of $V_{tip}$~0.5 V, labeled A, B and C. For comparison, the curve is superposed with a capacitance curve acquired with a micron-size gate in place of the tip. Supplementary Fig. 1 shows the behavior over a larger voltage range. To help explain the physical origin of the broad peaks, we first examine fine-structure peaks that also appear in the data, as shown in Fig. 2b. The two curves in the figure were acquired under identical conditions but with a time delay of nine hours. We see that most, but not all, of the peaks are reproduced. Fig. 2c shows three curves acquired at the voltage marked by the red arrow in Fig. 2b, with the average shown to the right. The data are consistent in magnitude and peak shape with the resonance expected for single-electron tunneling[12]. Supplementary Fig. 2 shows more details of the comparison between fine-structure peaks and the single-electron curve.

Given that we have identified the fine-structure peaks with individual electrons entering the donor layer, a natural explanation for the broader peaks, *A*, *B*, and *C*, is that they are formed by clusters of several electrons entering at nearly the same energy. If we convert capacitance to charge units, we find that peaks A and B each correspond to about 15 electrons entering the donor system; peak C is larger. What physics could gives rise to the broad resonances? It is possible that dense groupings of the donors result in electron puddles acting as small quantum dots[13]. In that scenario, an ensemble of puddles that have nearly the same addition energy spectrum could explain the peaks. However, given that the positions of the donors is random, as shown schematically in Fig. 1c, it seems unlikely that 15 such puddles would form within a radius of only 60 nm with sufficiently similar characteristics. Considering the opposite limit, a candidate for identical quantum objects is single silicon donors. However, according to Lieb's theorem, the maximum negative ionization for a molecule with K ions is Z+K-1, where Z is the total nuclear charge of the ions[5]. For a single donor this gives one, corresponding to the H$^-$ system. That would give only two peaks in a capacitance measurement, which is inconsistent with our observations (neglecting for the moment the perturbation of the tip). However, if we consider closely spaced Si donors effectively forming two-donor molecules (2DMs), then K=2, Z=2, this upper bound is relaxed and the theorem predicts that the system could actually bind up to five electrons (negative ionization of three).

To develop a model for a theoretical addition spectrum of 2DMs, our approach is to use the configuration-interaction method in the context of the effective mass theory[4,14-16]. In this approximation, each donor is regarded as a hydrogenic atom with an effective Bohr radius $a_0^* = 4\pi\varepsilon_0\kappa\hbar^2/m^*e^2$ and effective Rydberg energy $Ry^* = e^2/8\pi\varepsilon_0\kappa a_0^*$, where *e* is the electron charge, *m*\* is the electron effective mass, and $\kappa$ is the dielectric constant. In our sample, Si donors reside in Al$_{0.3}$Ga$_{0.7}$As, for which $a_0^*$=7.3 nm and $Ry^*$= 8.1 meV[17]. This energy scale is much greater than the thermal energy at our experimental temperature.

Before introducing the full model, we consider the likelihood of finding two donors in our sample spaced at a distance equal to or less than an effective Bohn radius, assuming a random distribution within the plane. The planar density of 1.25x10$^{16}$ m$^{-2}$ implies an average spacing of 8.9 nm, which is comparable to $a_0^*$. So we expect many donors will have a nearest neighbor sufficiently close to form a 2DM. This fraction must be balanced against the fraction of donors that will have more than one near-by neighbor. These more complicated molecules will have qualitatively different addition spectra. Supplementary Fig. 3 shows the statistical distributions of nearest neighbor distances. By integrating over the appropriate distribution, we find that 38% of the donors are expected to have zero or one nearest neighbor within $a_0^*$. This is the relevant percentage for our model, which considers the 2DM spectra based on the statistical distribution of distances between nearest neighbors $d$. The model does not account for clusters of three or more donors that we expect to pertain to 62% of the Si donors. However unlike isolated donors and 2DMs, clusters of three or more donors can have multiple geometric configurations. Hence, they are unlikely to form quantum objects with similar addition spectra which would sum to form the broad peaks.

Fig. 3a shows the configuration-interaction calculations of the 2DM electronic energies for all bound electrons as a function of separation $d$. The calculations include an image charge to approximate the potential applied by the tip, which tends to increase the electron confinement within the molecule. We see that each molecule can accommodate four electrons. This is not surprising given that an isolated H-atom, and a single Si donor, can accommodate two electrons. However, over the range of separations shown in Fig. 3a, without including the approximate tip potential the calculations predict that each 2DM would only hold two electrons (see Supplementary Fig. 4). Supplementary Fig. 5 shows more details of the tip confinement potential.

Fig. 3b shows schematically the full model we have developed. We consider a two-dimensional area of $\pi(60nm)^2$ with a fixed number of donors, labeled $i$ in Fig. 3b. We position the donors randomly within the area and find the nearest neighbor for each one. Each of these 2DM's (labeled $k$) has assigned to it an addition spectrum $\varepsilon^k_{1,2,3,4}$ based on the separation of the two atoms, as given by the configuration-interaction calculations (Fig. 3a). We account for the effects of non-nearest neighbors $i$ by adding the Coulomb energy $U^k_i$ to every quantum level due to all the other donors. The variations of this energy (each 2DM has a different configuration of neighbors) is the main source of broadening of the addition spectrum peaks. Understanding the detailed shape of each peak is subtle, stemming from the fact that the ionization of the system changes during the measurement; i.e., the donors become neutralized as charge is added to the layer. Our modeling shows that the resulting peak width is roughly 1 Ry*; the methods section presents more details of this part of the analysis. Of course, our model should only apply to about 38% of the donors, or roughly 1/3. We account for this in a simple way: instead of considering 140 donors, the nominal number in the probed area of $\pi(60nm)^2$, the model considers only 140/3 ~46. This is justified because donors with more than one neighbor within $a_0^*$ will on average have deeper energy levels; hence they will tend to be neutralized at relatively low voltages compared to the 2DMs. The model also includes the screening effect of the nearby 2D layer in our sample by using appropriately-positioned image charges. Lastly, to generate smooth curves to which we can compare our measurements, we perform the calculation for hundreds of random ensembles and average the results.

Fig. 3c compares the full 2DM model to the measured broad peaks with the background capacitance subtracted. The measured data are shifted horizontally to align peak C to the tallest peak predicted by the model. This is consistent with peak C lying near zero effective Rydbergs, the energy above which the electrons are unbound. No other free parameters were employed. We see that the model generally agrees well with the measurement, although some features in the data are not accounted for, such as the small peak near -7 Ry* and the relative sharpness of peak A. As indicated by the arrows, the model predicts that the peaks due to the third and four electrons ($\varepsilon_{3,4}$) will be unresolved. This is consistent with capacitance curves we have acquired in the presence of magnetic field which show a splitting of peak C; this effect is the subject of further investigations.

In conclusion, we have measured the electron addition spectrum of silicon donors using a scanning probe technique. The data are compared to a model which considers donor molecules effectively formed by nearest-neighbor silicon atoms. The overall agreement with the measurements suggests that this model captures the key physics of the system. The analysis highlights the high sensitivity of the molecules to external potentials. In particular, the data and modeling suggest that the voltage applied by the tip does not shift the addition spectrum in simple way; instead the quantum states are altered qualitatively as the number of electrons bound to a molecule increases from two to four. Future experiments include probing more dilute dopant systems and systems for which the lateral position of dopants can be controlled.

## ACKNOWLEDGMENTS

We gratefully acknowledge helpful comments and advice from R. C. Ashoori, M. I. Dykman, B. Golding, S. D. Mahanti and G. A. Steele. This work was supported by the Michigan State Institute for Quantum Sciences and the Nation Science Foundation, DMR-0305461.

**METHODS**

For the local probe measurements, we begin each data run by scanning the tip in both tunneling and capacitance modes to check that the surface is sufficiently clean and free of major electronic defects[11]. To acquire the capacitance curves, we position the tip about 1 nm from the GaAs surface and hold it at the fixed location while sweeping the tip voltage. To compensate for vibrations and drift effects, several curves are averaged together to achieve an acceptable signal-to-noise ratio.

For all the data presented here, the charging signal showed negligible phase shifts and hence can be considered as purely capacitance. Our sensor circuit includes a bridge that allows us to subtract away the background mutual capacitance of the tip and sample, ~20 fF. Hence our plotted signal represents the change in capacitance as a function of voltage. All voltages are plotted with respect to the effective zero voltage. This is the voltage for which no electric field terminates on the top electrode (gate or tip); it is shifted from the applied voltage by an amount equal to the contact potential, $V_{contact}$. For the PtIr tip used in the local probe measurements, $V_{contact} = 0.60$ V, as determined from *in situ* Kelvin Probe measurements[10]. For the gated capacitance data, the observed shift in the curves imply $V_{contact} = 0.12$ V; this value agrees reasonably well with the reported work functions of Ti and Au, in comparison to Pt and Ir[18].

In general, single electrons can be resolved by capacitance techniques at helium temperatures if the energy spacing to add successive electrons is on the millivolt scale or greater. As described in detail in reference 19, by measuring the capacitance $C$, we can detect charges entering the quantum system below the probe. We define the addition energy $\varepsilon_n$ as the energy for which the $n^{th}$ electron enters the system. As $V_{tip}$ increases from zero, the energy of an electron at the layer decreases as $-\alpha_{tip} e V_{tip}$, where $e$ is the magnitude of the electron charge and $\alpha_{tip}$ is the geometry-dependent proportionality constant, sometimes referred to as the voltage lever arm. In other words, electrons in the underlying 2D electrode are pulled toward the donor layer. The first electron will enter when the chemical potential equals the ground state energy of the one-electron quantum state, $\varepsilon_1 = E(1)$. As $V_{tip}$ increases further, the second electron will be induced to enter when the chemical potential equals the energy difference between two-electron and one-electron ground states, $\varepsilon_2 = E(2) - E(1)$. In general, $\varepsilon_n = E(n) - E(n-1)$, where we define $E(0) = 0$. The capacitance $C$ is given by $C \equiv dq_{tip}/dV \propto \partial <n>/\partial \mu$, where $dV$ corresponds to the excitation voltage, $\mu$ is the chemical potential, and $<n>$ is the expectation value for the number of electrons in the system.

For our experiment, there are two lever-arm parameters. For the gated measurements shown in Fig. 2a, the parallel-plate geometry and sample growth parameters give a proportionality constant of 1/4.0 with respect to the donor layer. For the local probe measurements, the relative scale factor between the two voltage ranges used in Fig. 2a imply $\alpha_{tip} = 1/10.8$. This is a reasonable value consistent with the expected tip-sample mutual capacitance[20-22].

With regard to our donor molecule model, we consider a two-dimensional area with donors $i$ positioned randomly. Each donor is paired with its nearest neighbor to form molecule $k$. (For a small fraction of donors, ambiguous cases can arise such as a equidistant nearest neighbors. In these cases the assignment of pairs is arbitrary.) To accurately simulate the capacitance measurement, we must consider that the ionization of the system changes throughout the measurement, as shown schematically in Fig. 3b. For example, for the first electron to enter the area, we assume all the donors are ionized. Hence we calculate the Coulomb shifts for each pair $k$, $\sum_i U_i^k$, using a charge of $+e$ for all donors. In this case, the pair that has the lowest energy $\varepsilon_1^k + \sum_i U_i^k$ receives the electron, filling its first state and thus contributing to the capacitance at this energy. For all subsequent electron additions into other pairs, we must consider that this particular pair no longer has two fully ionized atoms. In other words, for the second electron, which would likely enter some other pair, the Coulomb shifts will be slightly reduced due to the previous charge that has already entered the system and partially neutralized one pair of atoms. For simplicity, our modeling routine assumes perfect screening: every time an electron enters a 2DM, we add $-e/2$ to each atom of the pair. To generate the characteristic capacitance versus voltage curve, the procedure is repeated for hundreds of random ensembles and the results are averaged.

      The result of this procedure is the addition resonances $\varepsilon$ are broadened ~1Ry*, with each peak taking a distinct shape, as shown in Fig. 3c. The reason the model gives only three peaks despite the fact that there are four electrons per molecule is simple: both $\varepsilon_3 = E(3)-E(2)$ and $\varepsilon_4 = E(4)-E(3)$ are less than 1 Ry*, hence they are not resolvable as individual peaks. The reason the peaks have distinct shapes is more subtle, arising from the ionization effects described above. For example, for the second electron additions $\varepsilon_2$, on average there are fewer ionized charges in the donor layer than for the first electron additions $\varepsilon_1$. Therefore the overall Coulomb shift is reduced for the second electrons, which form peak B, as well as the broadening effect due to the randomness in the donor positions. For this reason the model predicts that peak B will be sharper than peak A. The shape of peak C is also broadened by the proximity of $\varepsilon_3$ and $\varepsilon_4$.

**FIGURES**

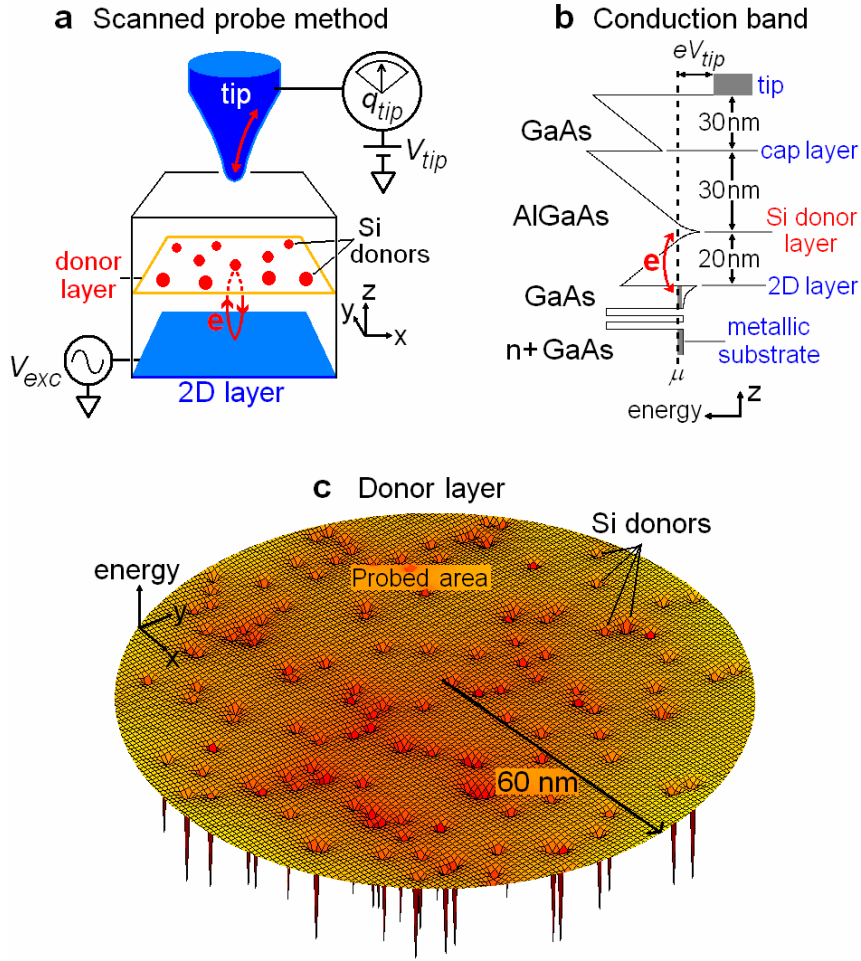

**Figure 1 | Capacitance-based scanning probe technique to detect donor charging. a,** Schematic of the key layers in the gallium-arsenide [001] heterostructure sample and the measurement technique. An excitation voltage can cause charge to resonate between the Si donor layer and a two-dimensional (2D) layer, which represents an ideal base electrode. This results in image charge appearing on a sharp conducting PtIr tip. A cryogenic transistor attached directly to the tip is used to measure the charging. The donor layer consists of silicon atoms confined to a single monolayer with respect to the *z* direction, but randomly positioned with respect to the *x-y* direction with an average density of $1.25 \times 10^{16}$ m$^{-2}$. At zero applied voltage, at least 90% of the Si atoms are ionized (i.e., donated an electron), as discussed in Supplementary Figure 1. Magneto-capacitance measurements conducted in the kHz frequency range indicate negligible donor-layer conductivity for identical samples cut from the same wafer. **b,** More detailed conduction band diagram of the sample. The excitation voltage is applied to a degenerately doped substrate that acts as a metallic electrode. Above this is the 2D electron layer. It is separated from the metallic substrate by a superlattice tunneling barrier; the tunneling rate into the 2D layer is an order of magnitude greater than the 20 kHz excitation frequency we employed. Hence for this experiment, the 2D layer can be regarded as being in ohmic contact with the substrate. **c,** Schematic of the area probed by the technique with Si donors represented as hydrogenic potentials. For our experimental geometry, the radius of the area over which we are probing is set mostly by the tip-donor layer separation, which is approximately 60 nm[20,21]. Within this area, on average we expect 140 donors.

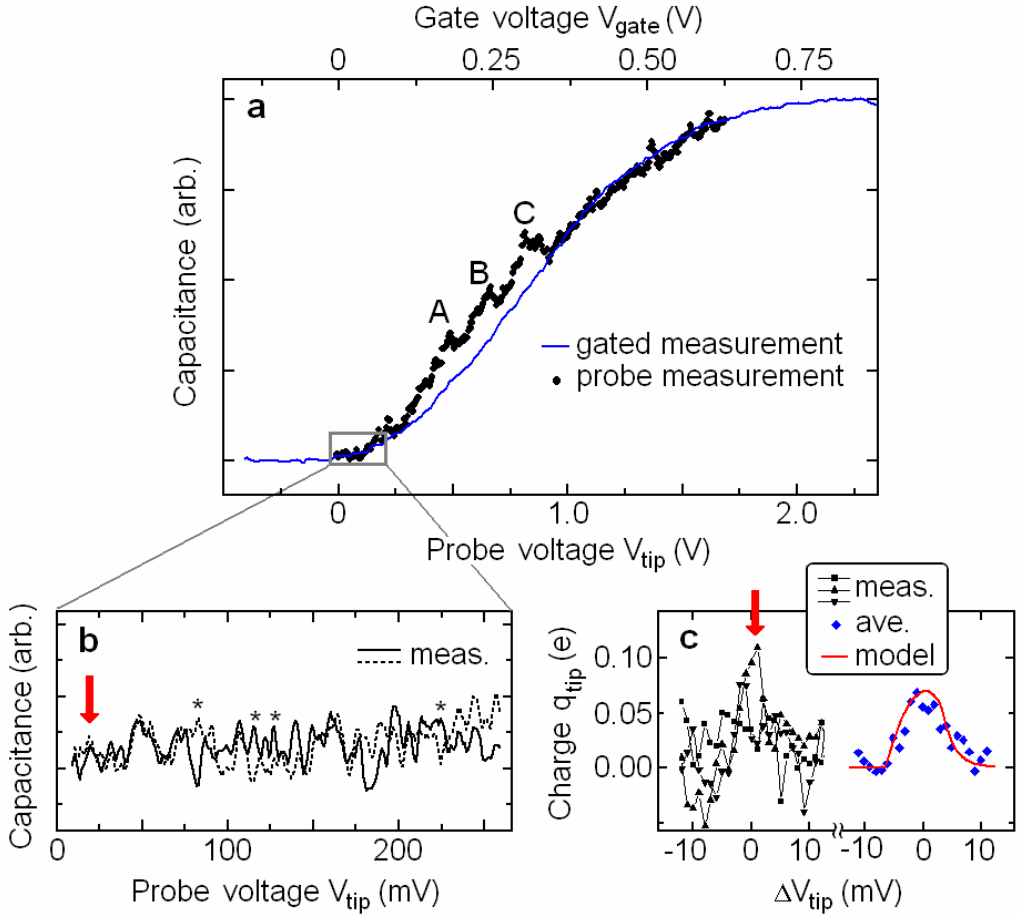

**Figure 2 | Representative capacitance data. a,** Capacitance measured with our local probe superposed with a curve acquired with a micron-size gate in place of the tip. The curves have different voltage ranges and the vertical scale of the probe measurement is exaggerated greatly relative to the gated measurement, consistent with differences in the probed area and with distinct lever-arm parameters for the two measurements (see methods section). The probe measurements consistently showed three broad peaks labeled A, B and C, whereas the gated measurement effectively provides a background curve, as discussed in Supplementary Fig. 1. To show clearly the characteristic structure, for the probe measurement we show the average of three measurements acquired at different locations. For both the local and gated curves, the voltage scales are plotted relative to the effective zero voltage, compensating for the contact potentials between the materials. The excitation voltage amplitude was $V_{exc}$=15mV rms for both curves. See Supplementary Fig.1 for more details. **b,** Capacitance versus tip voltage curve over the expanded voltage range, as indicated. To investigate the structure in detail, these data were acquired with smaller excitation amplitude of 3.8 mV rms. The two curves were acquired under identical condition at the same location, but with a time delay of nine hours. Much of the fine structure is reproduced, with asterisks marking missing or shifted peaks. These changes likely reflect long time scale charging and discharging of DX centers[16]. **c,** Three curves acquired at the voltage marked by the red arrow in **b** (same location) with an excitation voltage of 3.8 mV. The vertical scale has been converted to charge units $q_{tip}$. The plot to the right shows the average of the three measured curves, compared to a model curve which shows the semi-elliptical peak shape expected for single-electron tunneling[12]; see Supplementary Fig. 2 for a more detailed description.

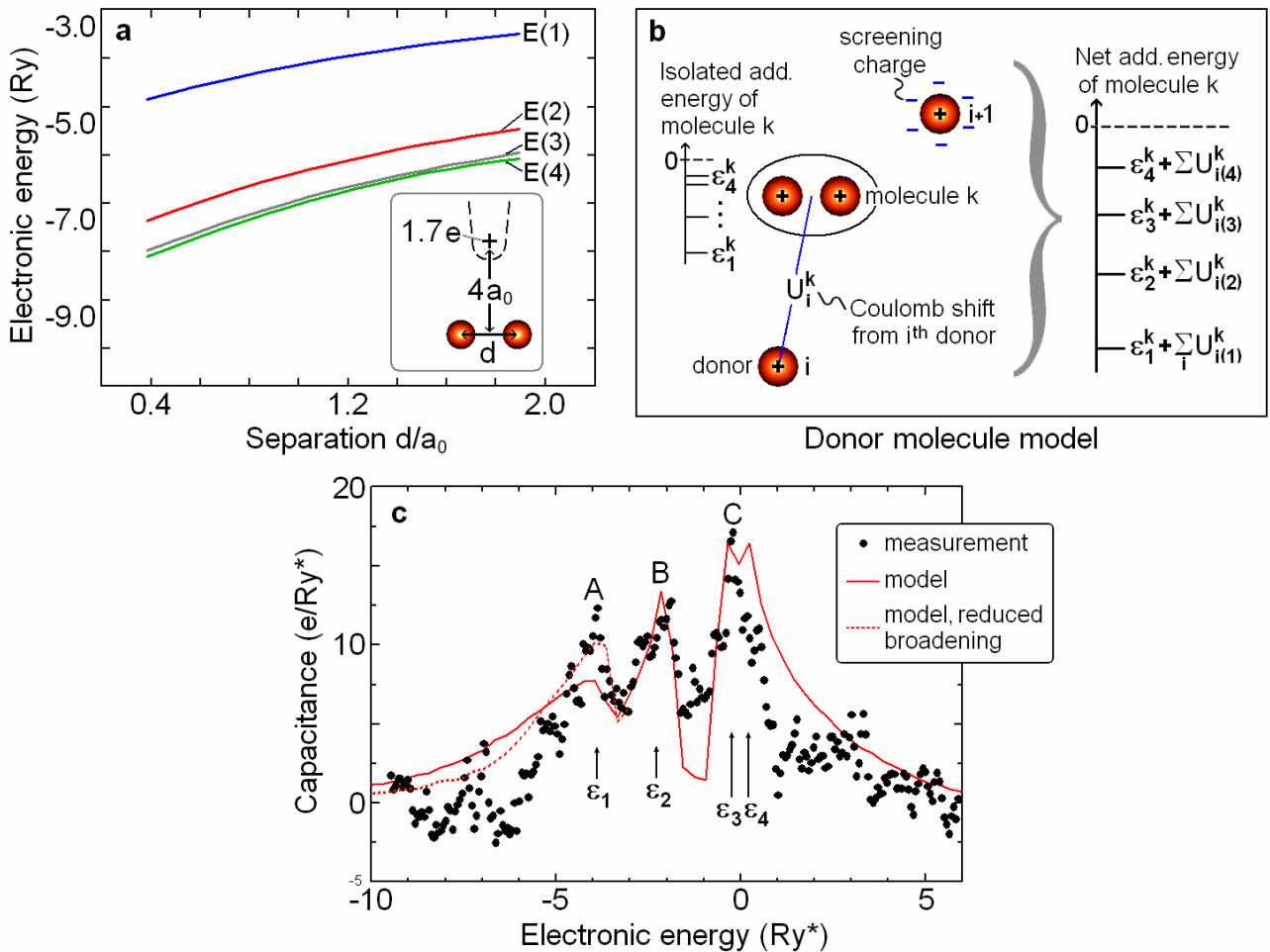

**Figure 3 | Two-donor molecule theory compared to experiment. a,** Configuration-interaction calculations of the 2DM electronic energies for all bound electrons as a function of separation of the two ions $d$. The calculations include an image charge to approximate the potential applied by the tip (inset). The model predicts four bound electrons for each donor molecule. Including the approximate tip potential is key, as without the extra confinement the calculations show only two bound electrons, as shown in supplementary Fig. 4. **b,** Schematic representation of the full modeling procedure. We start from a random ensembles of donors and group them into nearest-neighbor pairs to form molecules $k$. The electronic energies shown in **a** are used to calculate the isolated addition energy of each molecule, $\varepsilon_1^k=E(1)$, $\varepsilon_2^k=E(2)-E(1)$. Lastly, the model includes the Coulomb energy shift from all non-nearest neighbors; we account for the fact that this shift will be different for successive electrons due to changes in the screening charge of non-nearest neighbors donors, as described in the methods section. **c,** Comparison between experiment and theory. To allow a direct comparison, we subtract away the background capacitance from the measurements, plot the voltage in units of effective Rydbergs (scale factor $\alpha_{tip}$ / 8.1 x $10^{-3}$ V/Ry*), and plot the measured capacitance in units of electrons per Ry*. Although the match between experiment and theory is not exact, the overall agreement suggests that the donor-molecule model captures the correct physics. The dotted red curve addresses the discrepancy with regard to peak A, for which the predicted peak is significantly broader than the measurement; here we have reduced the broadening in the calculation by positioning the 2D layer 8 nm closer to the donor layer. In reality, reduced broadening may arises from increased screening in the donor layer.

# SUPPLEMENTARY FIGURES

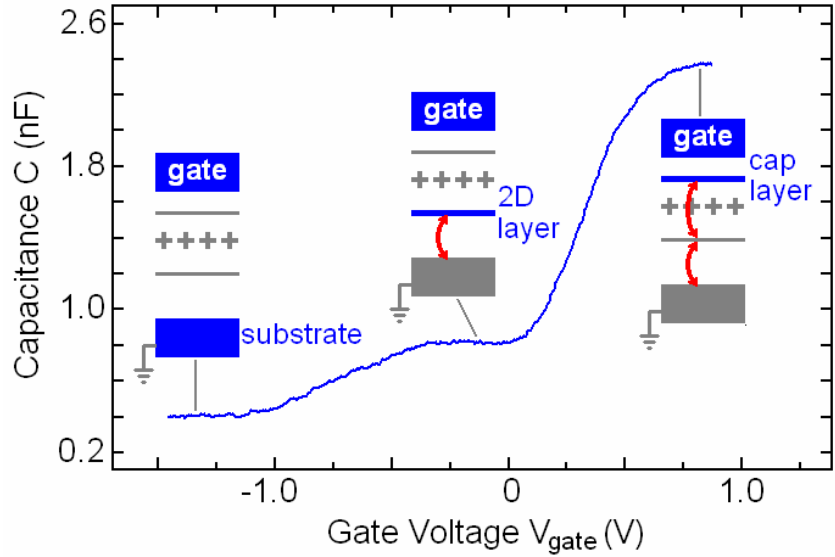

**Supplementary Figure 1 | Measurement of sample capacitance using a surface gate electrode.** To establish the baseline capacitance behavior in our system, we performed a measurement using a planar gold-titanium gate of area $5.7 \times 10^{-7}$ m$^2$ microfabricated onto the sample surface in place of the scanning probe tip. The sample was cut from the same wafer as the sample used in the local probe experiment. We see that the capacitance increases with gate voltage, forming three plateaus. The sketches show our interpretation of the plateaus as different layers below the gate accumulate charge (blue). In addition to the substrate electrode (170 nm below the surface) and 2D layer (80 nm below the surface), charge can also become trapped in the cap layer (30 nm below the surface). We do not display the data beyond 1.0 V, for which the signal begins to show a significant phase shift, indicating that charge is leaking directly onto the gate electrode. For this measurement, the gain of the capacitance signal was determined from sample geometry and gate area. For all other measurements, the signal was amplified with our HEMT sensor charge sensor, for which the gain was measured independently.

In addition to showing the accumulation of electrons in the 2D and cap layers, the gated-capacitance measurement allows us to estimate the density of ionized donors. This follows from the observation that the 2D electron system is fully formed at zero applied voltage. Of course, ionized donors introduce electric field; this in turn changes the slopes of the conduction band potential as shown in Fig. 1b. These slopes must be sufficiently steep to allow the conduction band to dip below the Fermi level at the 2D location. Solving Poisson's equation with this constraint yields a density of ions equal to at least 90% of the growth Si density of $1.25 \times 10^{16}$ m$^{-2}$. Hence, most of the Si atoms have indeed donated an electron and are ionized at zero applied potential.

Between the 2D layer and cap layer plateaus, there are small hints of structure. But unlike the local probe measurements, there are no clear peaks consistent with charge entering the donor layer. The reason such peaks are not resolved, both here and in previous capacitance studies, probably arises from the larger area probed in gated measurements. Micron-size areas are more likely to contain at least one severe defect or impurity that allows charge to enter higher layers in the sample (including surface states) without interacting directly with the donor system. In contrast, the local measurements presented in this paper probe an area of $2.3 \times 10^{-14}$ m$^2$. This is two or more orders of magnitude smaller than gated measurements, making it much less likely to find such a defect. The gated measurement can be regarded as providing the background plateau structure for measurements with the localized probe, which can be interpreted as a superposition of the plateau structure and donor layer peaks.

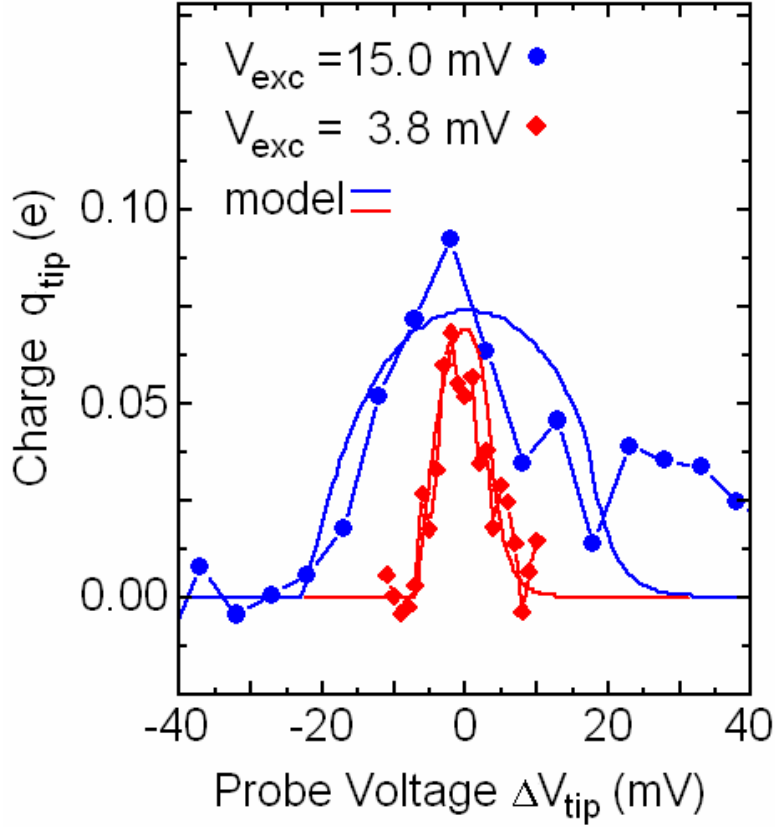

**Supplementary Figure 2 | Comparison of fine peaks to single-electron model.** The blue dots show a fine-structure peak measured using $V_{exc}$=15.0 mV and the red diamonds show a fine-structure peak measured using $V_{exc}$=3.8 mV (the same data shown in Fig 2c). The data were acquired in different tip locations; in both cases the peaks were selected as they are relatively well-isolated from neighboring peaks. We have converted the vertical capacitance scale to show the rms charge induced on the tip in units of *e*, as was done for Fig. 2c. The data are compared to two model curves which show the expected semi-elliptical peak shapes for single-electron charging for each measurement[12].

With regard to the widths of the peaks, in the low-temperature limit, the width of the model curves is set by the excitation amplitudes. We include in the model additional broadening due to the output filter of our lockin amplifier. This gives the asymmetry to the model peaks. We see that the single-electron model agrees reasonably well with the measurements. With regard to the vertical scale, if all the electric field lines were captured by the tip, the magnitude of model single-electron peaks would be 0.99 e and 0.92 e, respectively. However, to achieve a good fit, the heights of the model curves are scaled by 0.075. This peak height is roughly consistent with expected captured electric flux for single-electron charging within the donor layer[22], for which the scale factor should be approximately $\alpha_{tip}$=1/10.8=0.093. Hence, we conclude that the isolated fine-structure peaks likely reflect individual electrons entering the donor layer below the tip.

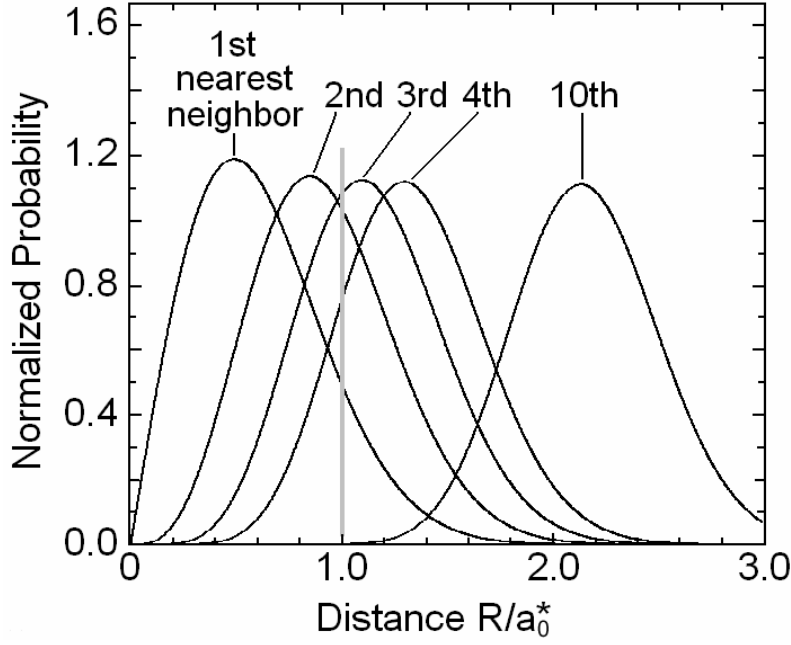

**Supplementary Figure 3 | Statistical nearest-neighbor distances for donors in our sample.** To gauge the likelihood of finding 2DMs in our system, we show the statistical nearest-neighbor distances for donors dispersed randomly within a two-dimensional layer. Nearest neighbor distances essentially follow Poissonian distributions. Selecting a donor at random, the probability to find its $m^{th}$ nearest neighbor between a distance $R$ and $R+dR$ is

$$\frac{(\pi R^2 \rho)^{m-1}}{(m-1)!} 2\pi R \rho \exp(-\pi R^2 \rho) dR ,$$

where $\rho$ is the two-dimensional density. For the curves shown here, we use the nominal planar donor density of Si in our sample, $\rho = 1.25 \times 10^{16}$ m$^{-2}$. The distances are given with respect to the effective Bohr radius of $a_0^* = 7.3$ nm. We see that we have a relatively dense donor layer. For example, by integrating the $1^{st}$ nearest-neighbor curve, we find that 88% of the donors have their first nearest neighbor within $a_0^*$; similarly, 62% of donors have $2^{nd}$ nearest neighbor within $a_0^*$. The relevant fraction with respect to our 2DM model is the percentage that have zero or one nearest neighbor within $a_0^*$. This is given by 1-0.62=0.38, or 38%.

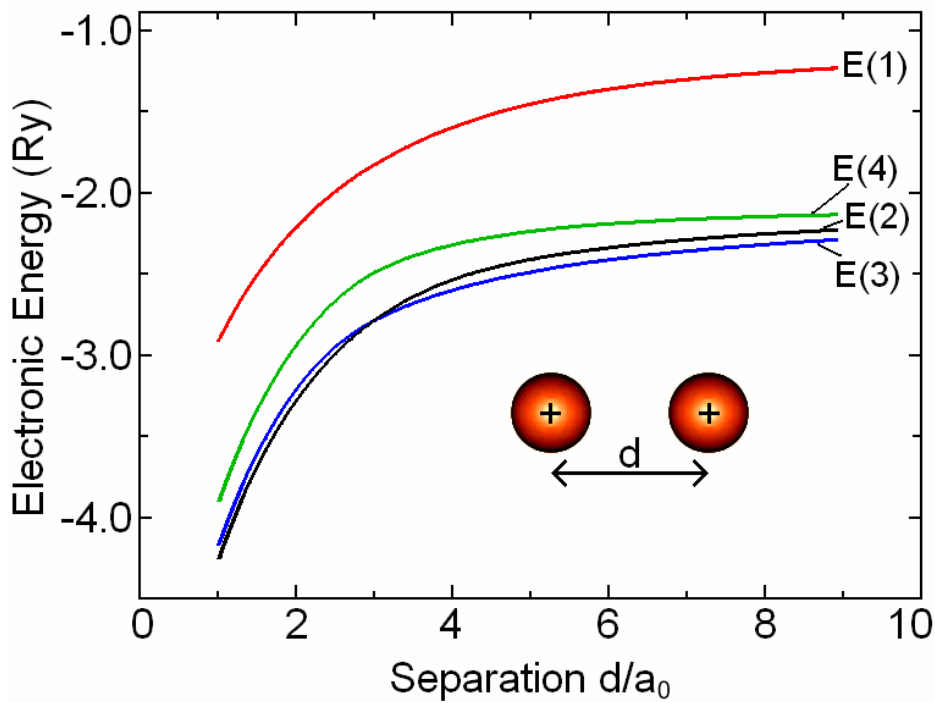

**Supplementary Figure 4 | Configuration-interaction calculation of the electronic energies of two-donor molecule.** Here we show the results of the simplest 2DM model where we have calculated the electronic energies for the first four electrons of two hydrogen nuclei (or Si donors in the effective mass approximation), separated by a distance *d*, but otherwise isolated, as shown in the inset. Interestingly, *E(3)* is lower than *E(2)* for large separations, but the two lines cross at ~$3a_0$. This means that at large separations, the molecule will hold three electrons, similar to the H⁻ state; but for small separations only two electrons can be accommodated. The intuitive picture is that the neutral system can polarize and weakly bind the third electron. However, this is prohibited for small separations for which the direct Coulomb repulsion dominates. We see that *E(4)* is always higher energy than *E(3)*, hence the fourth electron is never bound.

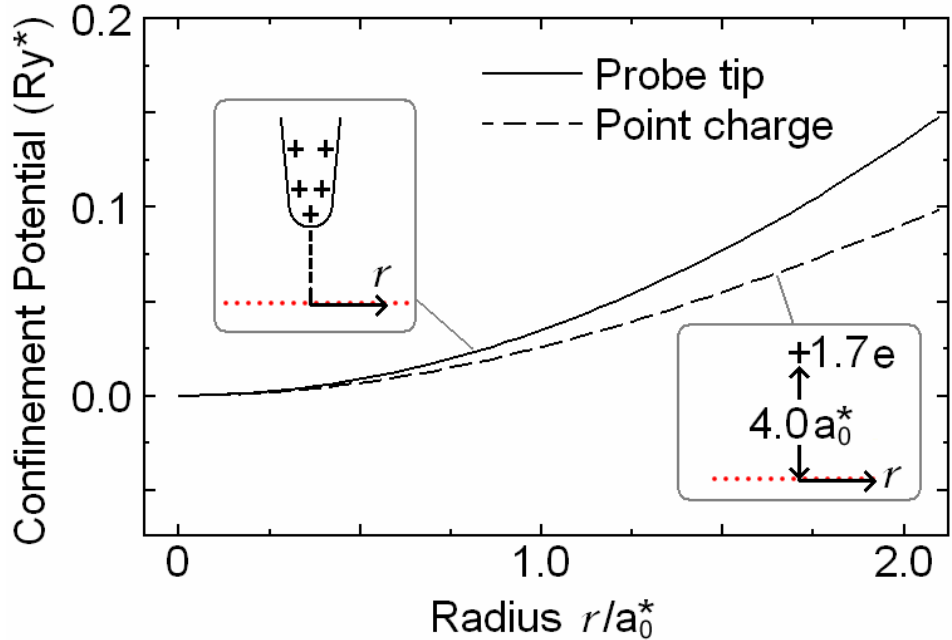

**Supplementary Figure 5 | Approximation of lateral confinement due to the tip potential.** With respect to the donor-layer plane, the positive tip voltage gives a curved background potential that tends to increase the confinement. The solid curve of shows the expected confinement potential for $V_{tip}$=0.45 V which is the potential in the vicinity of peak A. This curve is based on the mutual capacitance function given in reference 22 between a layer 60 nm below a dielectric surface and an arbitrarily narrow tip; $r$=0 is the point in the layer directly below the tip as shown in the inset. This function is somewhat broader near $r$=0 than similar curves in references 20 and 21. Hence, it represents a conservative estimate of the tip's confinement. We include this effect in our model by incorporating an image charge in the configuration-interaction calculations of 1.7$e$ at a distance of 4$a_0$, as indicated. The charge adds the confinement shown by the dashed curve; however, the basis set of functions for the calculation include no orbitals localized on this charge[15].

We see the image charge approximation is somewhat weaker than the expected confinement; moreover we take this potential as fixed, even though the tip voltage varies during the measurement. Hence, this is a very rough approximation of the tip's influence, necessitated by the computationally-intensive nature of the calculations. The decision to err on the side of weak confinement is justified by the fact that the confinement effect weakens for donors not directly below the tip.

Fig. 3a shows the corresponding 2DM calculations for the electronic energies for the first four electrons. In this case we see that $E(3)$ is lower than $E(2)$ even for small separations. Hence this spectrum shows that the third electron will always be bound. Moreover, the forth electron is also bound, but very weakly. All subsequent electrons are unbound in this calculation.